# Ultrahigh pyroelectricity in monoelemental 2D tellurium

Hari Krishna Mishra[1], Ayushi Jain[1], Dalip Saini[1], Bidya Mondal[1], Chandan Bera[1,*], Shanker Ram[2,*] and Dipankar Mandal[1,*]

[1]*Quantum Materials and Devices Unit, Institute of Nano Science and Technology, Knowledge City, Sector-81, Mohali 140306, India*

[2]*Materials Science Centre, Indian Institute of Technology, Kharagpur–721302, India*

We report an ultrahigh pyroelectric response in van der Waals bonded layers of two-dimensional (2D) tellurium (Te) nanosheets (thickness, d = 4 to 5 nm) at periodic on-off temperature oscillations. For the first time a large pyroelectric coefficient, $P_c \sim 3\times10^3$ µC.m$^{-2}$.K$^{-1}$, is observed which is eightfold higher than the traditional state-of-the-art pyroelectrics (lead zirconate titanate, PZT). The first-principles calculations point out that the breakdown of centro-symmetry in the 1-3 Te-layers (P-3m1 space group) of a non-centrosymmetry (higher-order symmetry of C2 space group) on an angular twist in the Te-Te bonds of an exotic electronic state in 2D Te. The angular Te-Te twisting elicits a surface-enhanced Raman band at 101 cm$^{-1}$ (absent in bulk Te). The stimulation of the Born effective charge, in-plane piezoelectricity and thermal expansion coefficient are shown to tailor the large pyroelectricity. Thus, 2D Te nanosheets present a new paradigm for the wide application of pyroelectric materials for developing thermal energy-based flexible electronics.

Pyroelectricity refers to the conversion of thermal energy into electrical energy due to the change in switchable spontaneous polarization ($P_s$) within polar dielectrics at a temporal temperature oscillation. This phenomenon offers a promising approach for thermal energy harvesting, allowing the utilization of otherwise wasted heat [1–3]. In this scenario, a pyroelectric material is employed to convert thermal energy into usable electricity as its $P_s$ is varied in stipulated conditions. The temperature sensitivity of the pyroelectrics is critical for designing detectors, neuromorphic computing, energy storage devices, thermal energy harvesters and many other devices [1-5]. Three-dimensional pyroelectrics have been well-explored for pyroelectric energy harvesting, however, their figure-of-merits (FoM) are limited by their intrinsic properties such as high dielectric permittivity and less Born effective charges (BECs) [3]. Despite the well-known 2D materials for their piezo- and ferro-electric properties, a comprehensive understanding of their pyroelectric properties has not been explored in depth to date [3-5]. Recently, Liu and Pantelides studied in-depth mechanisms of pyroelectricity in 3D and 2D chalcogenides by the first-principles calculations [3]. According to the modern theory of polarization of electron cloud distribution and Born's theory of quantum-mechanical electron-phonon (e-p) renormalization, the interactions of electrons with the lattice vibrations can lead to giant pyroelectricity [3,5]. When a pyroelectric material undergoes temperature (T) oscillations, $\partial T/\partial t > 0$ over a time 't' scale, it is strained, which leads to the displacement of the atoms from their equilibrium positions. Consequently, an 'e-p' interactions promote the net polarization [4,5]. The pyroelectric coefficient ($P_c$) and FoMs are two basic engineering parameters to elucidate the selection of the materials as an efficient pyroelectric nanogenerator (PyNG) [6-10]. The net $P_c$ at a mechanical stress (σ) is expressed in the Born-Szigeti theory of pyroelectricity as [3],

$$P_c = \left(\frac{dP_s}{dT}\right)_\sigma = P_{c1} + P_{c2} = \left(\frac{\partial P_s}{\partial T}\right)_\varepsilon + \sum_i \left(\frac{\partial P_s}{\partial \varepsilon_i}\right)_T \left(\frac{d\varepsilon_i}{dT}\right)_\sigma \tag{1}$$

where $P_{c1}$ and $P_{c2}$ refer to the primary (at constant strain, $\varepsilon$) and secondary part of pyroelectricity as per thermal expansion and piezoelectric coefficient. [6]

The monoelemental 2D chalcogenides possess excellent piezo-, pyro-, and ferro-electricity [9,11], although their out-of-plane piezoelectric response is relatively small. Recently, exotic–electro-mechanical properties are observed in atomically thin α-Te films with an out-of-plane piezoelectric coefficient $d_{33}$ ~ 1 pm/V [12]. In this scenario, Te atoms order in 1D helical chains bonded via weak van der Waals forces along the c-axis at a few Te layers [13,14]. The large $P_s$ in traditional oxides induce a larger piezoelectric response than in polymers [6,8], but they are not mechanically flexible [8,9]. Therefore, monoelemental chalcogenides are a promising alternative of piezoelectric and pyroelectric oxides [11-13], which can better perform than traditional materials [9,14].

In this article, we report a pyroelectric response of the van der Waals bonded 2D Te layers of a flexible pyroelectric nanogenerator (PyNG) at room temperature. A large $P_c$ ~ $3 \times 10^3$ μC·m$^{-2}$·K$^{-1}$ is observed in a non-centrosymmetric 2D α-Te. The first-principles calculations with a density functional theory (DFT) are explored to optimize the α-Te crystal structure at 1 to 3 layers. The results are described with lattice images, diffraction patterns, and phonons of 2D Te at few layers.

Thin Te layers were exfoliated by sonicating Te slurry in isopropanol (IPA) followed by centrifuging, which is a simple method to exfoliate thin Te layers [15-17]. Their structure and topology were studied with X-ray diffraction (XRD), transmission electron microscopy (TEM), lattice images and selected area electron diffraction (SAED). The phonon dispersion, Raman bands and other characterizations confer the features of cohesive 2D Te layers.

The geometric α-Te structures optimized from the first-principles calculations with DFT in Figs. 1(a, b, c) illustrate the way in which Te atoms can order in mono- and tri-layers. The lattice parameters are varied in three types of lattices of the P-3m1 space group at the monolayer (trigonal crystal structure), while P-3m1 and C2 space groups (monoclinic phase) at the trilayer. A monolayer shows a centrosymmetry at the P-3m1 space group that is not energetically stable to form at a trilayer at the lowest ground state energy, $\Delta E = E_{Total}^{P-3m1} - E_{Total}^{C2} = 14.2$ meV. An energetically stable ($\Delta E \to 0$) α-Te trilayer thus orders in a non-centrosymmetry C2 space group depicted in Fig. 1(c). The internal bond angles ($\theta_1$ and $\theta_2$) are triggered by the van der Waals interaction between the atomic layers. The angle $\theta_1 = \theta_2 = 88.17°$ at a monolayer is twisted to $\theta_1 = 87.78°$ and $\theta_2 = 88.10°$ at a trilayer of the C2 space group, which is a prime factor of exhibiting a large in-plane piezoelectricity. Non-equivalent Te–Te bonds and differences in $\theta_1$ and $\theta_2$ clearly display the absence of mirror planes. Thus, an induced in-plane piezoelectricity involves an effectively enhanced $P_{c2}$ in eq. (1). Furthermore, the ZA, TA and LA modes are shown in the phonon dispersion curves in Figs. 1 (d, e) for the monolayer and trilayer, respectively. The absence of imaginary modes at any high symmetry directions confirms both mono- and tri-layers α-Te are dynamically stable. The acoustic modes (e.g., ZA) are converged at Γ and M points in a duly 'lattice parameter (a)' expanded towards the trilayers.

Typical atomic force microscopy (AFM) images in Fig. 2(a) present the topography of 2D Te exfoliated sheets of w = 50-250 nm widths. When scanned at three different spots 1, 2, and 3, their average thickness 'd' is varied at a 5-6 nm scale, with a ± 1.2 nm deviation from image to image, as plotted in Fig. 2(b) along the surface. Image 3 of an orange-colored contrast, at a zoomed scale (right panel) in Fig. 2(a), shows it contains nanofibrils of L = 100-120 nm lengths (w = 20 - 25 nm). Nanofibrils are surrounding the intense image 2 of a cluster. The TEM images in Fig. 2(c) reveal round-shaped Te at the surface energy dominates in small crystallites (w = 4-10 nm) of $d_{102}$ = 0.2325 nm arrays. In Figs. 2(d, e), $2d_{100}$ = 0.7730 nm chains are stacking (a superlattice) onto the $d_{102}$ surface at an angle $\varphi = 51.0°$, which is reduced from $56.4°$ at $\tan\varphi = 2a/c \cong 1.5043$ in bulk α-Te [18]. A compressed '$\varphi$' reveals a duly compressed $2a/c = 1.235$ from 1.5089 shown in XRD of thin α-Te sheets. A SAED pattern taken form a sheet in Fig. 2(f) displays families of $d_{101} = 0.3210$ nm and $d_{102} = 0.2325$ nm lattice arrays of sharp spots, which describe a P3$_1$21 hexagonal lattice (which is derived from a long-range order symmetry of C2 space group) [18, 19]. Here, $\gamma = 58°$ is reduced (from an ideal 60° value) in a presumably strained (compressed) lattice in a relation [20],

$$\frac{1}{d^2_{hk\ell}} \cong \frac{4(h^2 + k^2 + hk)}{3a^2} + \frac{\ell^2}{c^2} \text{ , at } \gamma \to 60° \quad (2)$$

The XRD peaks are well illustrated at a closer view of the prominent (101) peak in Fig. 3(a) and its $d_{101}$ value is reduced by 0.71 % from 0.3231 nm in rather thick sheets (0.3234 nm in a α-Te bulk [18]) to at 0.3208 nm in the 2D Te layers. The average lattice parameters $a = 0.4435$ nm (0.4452 nm) and $c = 0.5905$ nm (0.5925 nm) determine a lattice volume $V_c \equiv a^2c\cdot\sin\gamma \cong 0.0985$ nm$^3$ (0.0996 nm$^3$) at $\gamma \sim 58°$ for the exfoliated (bulk) Te sheets from the Rietveld refined XRD patterns. An induced microstrain $\gamma_m = 0.39$ % (0.30% before exfoliation) is estimated from the Williamson-Hall (W-H) plot of the peak broadening [22]. Consistently, a larger $V_c \equiv 0.1020$ nm$^3$ is known in bulk α-Te of a crystal density $\rho_c = 6.23$ g/cm$^3$ [18], which is enhanced up to 3.53 % in the 2D Te sheets. The changed stimuli are shown in histograms in Fig. 3(b).

Three distinct Raman bands of phonons are observed in the bulk Te in Fig. 3(c) at 90, 119 and 139 cm$^{-1}$ from well-known $E_1$, $A_1$ and $E_2$ vibration modes of Te atoms, respectively, in a helical chains [23–25]. In thin 2D Te layers, the doublet $E_1/E_2$ degeneracy is lifted up so that the 90 cm$^{-1}$ band is split up into two parts of 84 cm$^{-1}$ and 92 cm$^{-1}$, while the 139 cm$^{-1}$ band is split up into those of 143 cm$^{-1}$ and 151 cm$^{-1}$. The 119 cm$^{-1}$ band is shifted at 120 cm$^{-1}$. In addition, a surface-enhanced Raman band emerged at 101 cm$^{-1}$ in the chain twists (B$_1$) along its length. As usual, the most intense band (A$_1$) 120 cm$^{-1}$ represents the symmetric oscillation of Te atoms in the chains. This is a so-called a 'chain breathing' vibration. The E$_1$ and E$_2$ describe the 'Te-Te-Te' bond bending and asymmetric bond stretching along the chains, respectively. The E$_1$ involves two types of 'Te-Te-Te' angle bending of transverse (TO) and longitudinal (LO) oscillations at the chains, which are separated apart at effectively thin 2D Te layers. In the E$_2$ mode, simply the 'Te-Te-Te' is rocking along the chain $\perp$ to the basal plane. A derived energy-level diagram in Fig. 3(d) describes a way for the 2D Te layer to regulate the phonons. In response to the surface energy, a 2D Te occupies the first excited electronic state $^3S_1$ above the $^1S_0$ bulk state. Thus, the $^3S_1$ state is thermo-sensitized to promptly render pyroelectricity due to phonon dynamics, as follows.

The phenomenon of pyroelectricity is characterized by the displacement vector ($\vec{D}$) of the charges at a temporal oscillation, $\partial T/\partial t \neq 0$, which can be expressed in an equation $\vec{D} = \epsilon_0 \vec{E} + \vec{P}$, where, $\vec{P}$, $\vec{E}$ and $\epsilon_0$ are the net polarization, electric field vector and dielectric permittivity of free space, respectively. For $\vec{P} = \vec{P}_{ind} + \vec{P}_s$, on an induced polarization $\vec{P}_{ind}$ in an applied electric field, we write $\vec{D} = \epsilon_0 \epsilon_r \vec{E} + \vec{P}_s$, where, $\epsilon_r$ is the dielectric permittivity of the material. In the absence of an electric field, $\vec{D} = \vec{P}_s$. Thus, $P_c$ at an electric field and a mechanical stress ($\sigma$) can be expressed ass [26,27],

$$P_c = \left(\frac{d\vec{D}}{dT}\right) = \left(\frac{\partial \vec{P}_s}{\partial T}\right)_{E,\sigma} \qquad (3)$$

To assess the pyroelectric response of the time-dependent temperature oscillations, a fabricated PyNG of 2D Te nanosheets is irradiated by an IR temperature. The output open circuit voltage ($V_{oc}$) and short circuit current ($I_{sc}$) are measured at the device when heated at $\Delta T \sim 5$ K (Fig. 4a). A (+) $I_{sc}$ pulse is observed at $\partial T/\partial t > 0$ and a (-) $I_{sc}$ peak at a convective cooling cycle. Thus, as depicted in Fig. 4(b, c), $V_{oc} \rightarrow 1$ V at $I_{sc} \rightarrow 13$ nA. In the Lang-Steckel method, an induced $I_{sc}$ proportional to '$\partial T/\partial t$' flows,

$$I_{sc} = P_c.A.(\partial T/\partial t) \qquad (4)$$

through area A in the circuit. In open circuit conditions, it is expressed as [28],

$$V_{oc} = \frac{P_c}{\epsilon}.d.\Delta T \qquad (5)$$

Here, $\epsilon$ is the dielectric permittivity of the sample of thickness d across the $P_s$ polarization direction. It implies $P_c \sim 3 \times 10^3$ µC.m$^{-2}$.K$^{-1}$ in the 2D Te nanosheets, namely 8 times larger than well-known pyroelectrics such as PZT [1,29], or 1.5 times than the PMN-0.25PT [1].

To realize that the giant pyroelectricity stems from an induced $P_s$ at the crystal unit cell scale, it is feasible to tune it at a heating-cooling cycle. A large $P_c$ observed in the 2D Te sheets is validated from the first principles calculations. Thus, a trilayer Te of a C2 space group shown in Fig. 1(c), having a unique polar axis and generating net electric dipole moment [30], exhibits in-plane spontaneous polarizations $P_{sp,a} = (-)1.89$ C/m$^2$ and $P_{sp,b} (-) 0.93$ C/m$^2$. These values are nearly $10^3$ times larger than reported for a MoSSe monolayer of $P_{sp} = (-) 14.15 \times 10^{-4}$ C/m$^2$ [31] and nearly 2 times than PZT of $P_{sp} = (-) 0.78$ C/m$^2$ [1]. The in-plane piezoelectric coefficient $e_{22} = 16.33 \times 10^{-10}$ C/m is five times enhanced relative to $3.74 \times 10^{-10}$ C/m for MoSSe [32]. To develop the underlying strategy, we studied the Born effective charges (BEC) that backup the $P_c$ in proportionality to the piezoelectric coefficient ($e_{ij}$), and the thermal expansion coefficient ($\alpha_a$) [2]. As shown in Table I (Fig. 4), an average BEC at 7$^{th}$, 8$^{th}$, and 9$^{th}$ atoms of a trilayer α-Te is nearly 3 times the values at Pb atoms and nearly 2 times the values for the Zr/Ti atoms in PZT ceramics.

The other Te atoms have higher BEC than the $O^{2-}$ (PZT). Thus, in-plane BECs are relatively higher than PZT [33]. Further, the estimated $\alpha_a$ = 15.94 K$^{-1}$ is far higher than MoSSe of 7.09 K$^{-1}$ [33]. The various estimated parameters are given in Table SII that promote the in-plane pyroelectricity triggered in 2D α-Te layers.

In particular, the in-plane geometry plays a vibrant role in tuning high pyroelectric response and excellent BEC, piezoelectricity, and $\alpha_a$ values. A possible mechanism of giant pyroelectricity in 2D Te-based in-plane PyNG. Further, FoMs are evaluated from $F_i$ = 0.59 nC.m.J$^{-1}$ and $F_v$ = 13 m$^2$.C$^{-1}$ as [34],

$$F_i = \frac{P_c}{C_e} = \frac{P_c}{\rho C_p}, \quad (6)$$

$$F_v = \frac{P_c}{C_e \varepsilon_{33}^\sigma} = \frac{P_c}{\rho C_p \varepsilon_{33}^\sigma} \quad (7)$$

with $C_p$ = 1.74 J g$^{-1}$ K$^{-1}$ specific heat, $\rho$ = 2.89 g/cm$^3$ density, $C_e$ = 5.03 J.cm$^{-3}$·K$^{-1}$ volume-specific heat and $\varepsilon_{33}^\sigma$ = 5 dielectric permittivity at given stress. As compared in Fig. 4(d), these are far higher pyroelectric stimuli than known so far [1,34].

In summary, a pyroelectric nanogenerator is fabricated from 2D Te sheets for thermal energy harvesting. It provides an unprecedented ultrahigh $P_c$ ~ 3×10$^3$ µC.m$^{-2}$.K$^{-1}$, and corresponding FoMs ($F_i$ = 0.59 nC.m.J$^{-1}$ and $F_v$ = 13 m$^2$/C) under periodic on-off temporal thermal oscillations. The profound properties arise from the centro-symmetry breakdown in exfoliated α-Te at a few atomic 2D layers, as conferred with the first principles calculations using the density functional theory. The tailored values of the spontaneous polarization, the Born effective charges, thermal expansion coefficient and in-plane piezoelectricity are 1 to 3 orders of magnitude higher than those of well-known traditional pyroelectrics. As a result, the fascinating pyroelectric response developed in the 2D Te sheets opens a new avenue for harvesting thermal energy and its technologies for the next generation of flexible electronics.

The authors gratefully acknowledge the financial support from the Science and Engineering Research Board, Department of Science and Technology, under the project (No. CRG/2020/004306), Government of India. A. J. is thankful to the University Grant Commission for awarding a fellowship (Grant No. 201610051125).

**Corresponding author**

Chandan Bera *(chandan@inst.ac.in),* Shanker Ram *(sram@matsc.iitkgp.ac.in)* and Dipankar Mandal *(dmandal@inst.ac.in).*

**114**, 436 (2015).

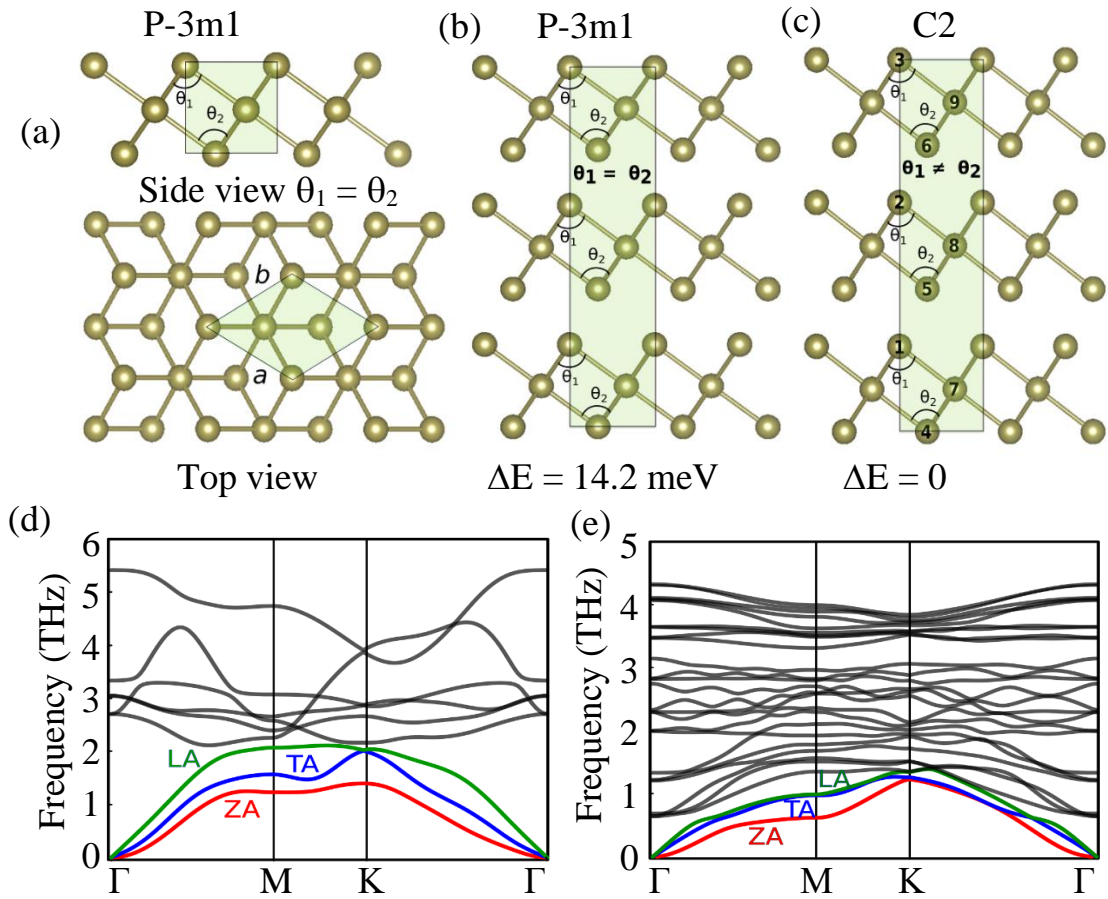

FIG. 1. Geometrical structures of 2D α-Te nanosheets; (a) a monolayer with side and top-views of a P-3m1 space group and (b, c) a trilayer with side-views of P-3m1 and C2 space groups (a unit cell at a shaded colour). Phonon dispersion α-Te curves of (d) mono and (e) tri-layers.

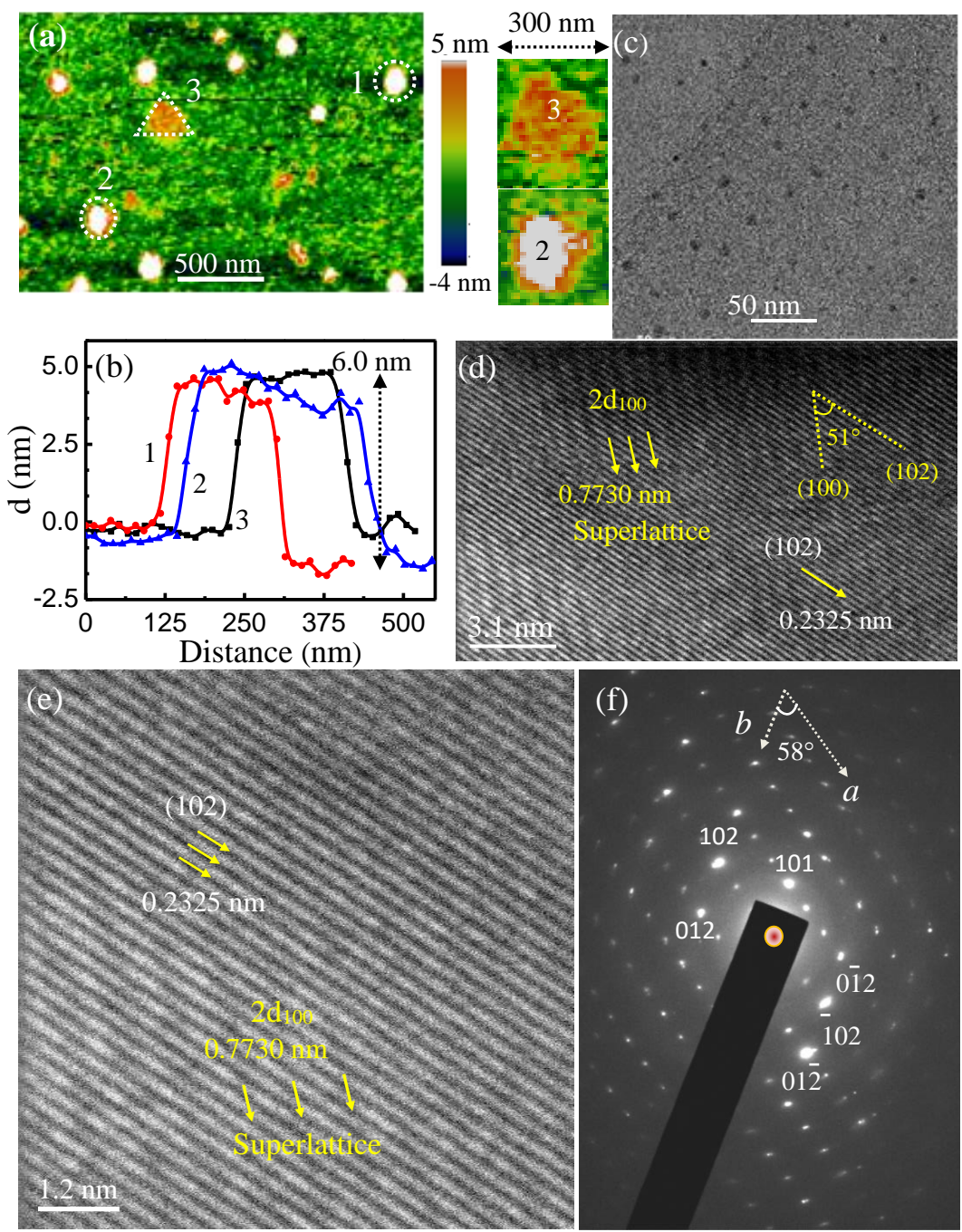

FIG. 2. (a) AFM surface topographies (a close-up of regions 2 and 3 in the right) of Te exfoliated as 2D nanosheets and (b) thickness profiles of the sheets measured at three different 1→ 3 regions. (c) TEM images and (d, e) lattice patterns, with (f) a SAED (region e), from a Te nanosheet.

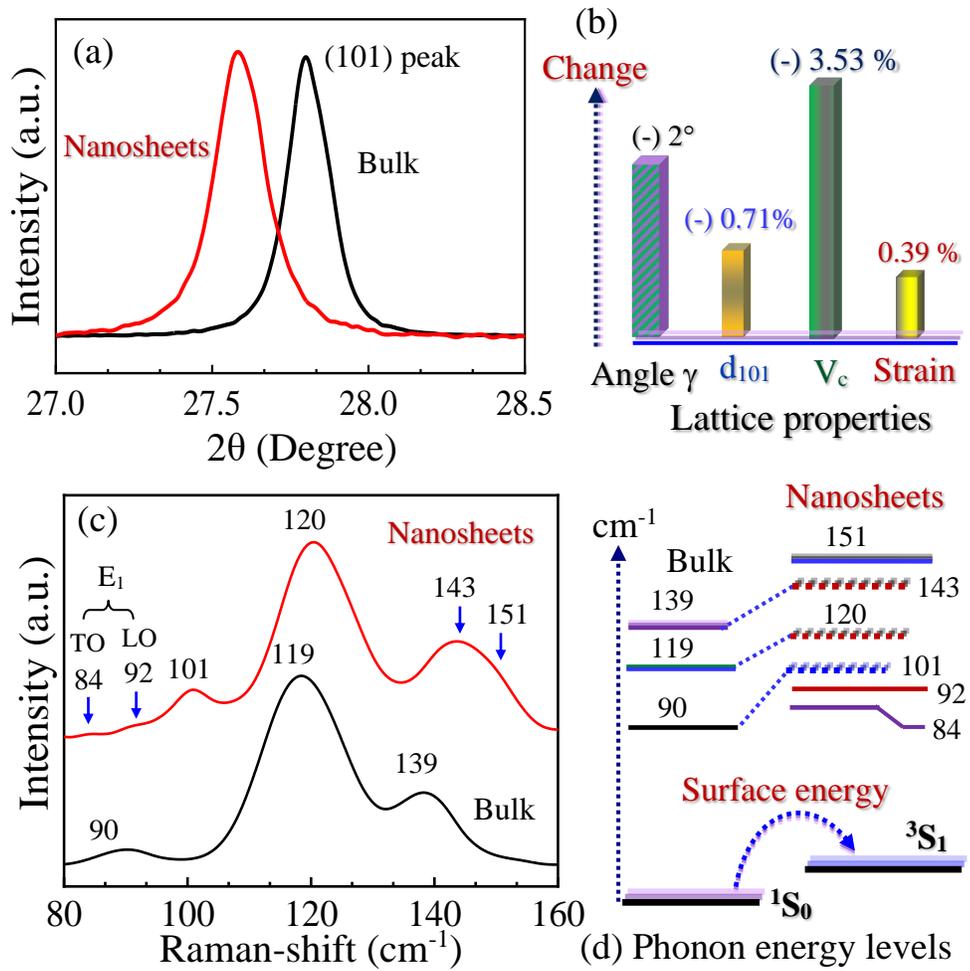

FIG. 3. (a) A prominent XRD (101) peak showing a marked blue shift in its $d_{101}$ value of an exfoliated α-Te from a bulk sheet, and duly changed (b) lattice properties, (c) Raman bands (phonons), and (d) phonon energy level.

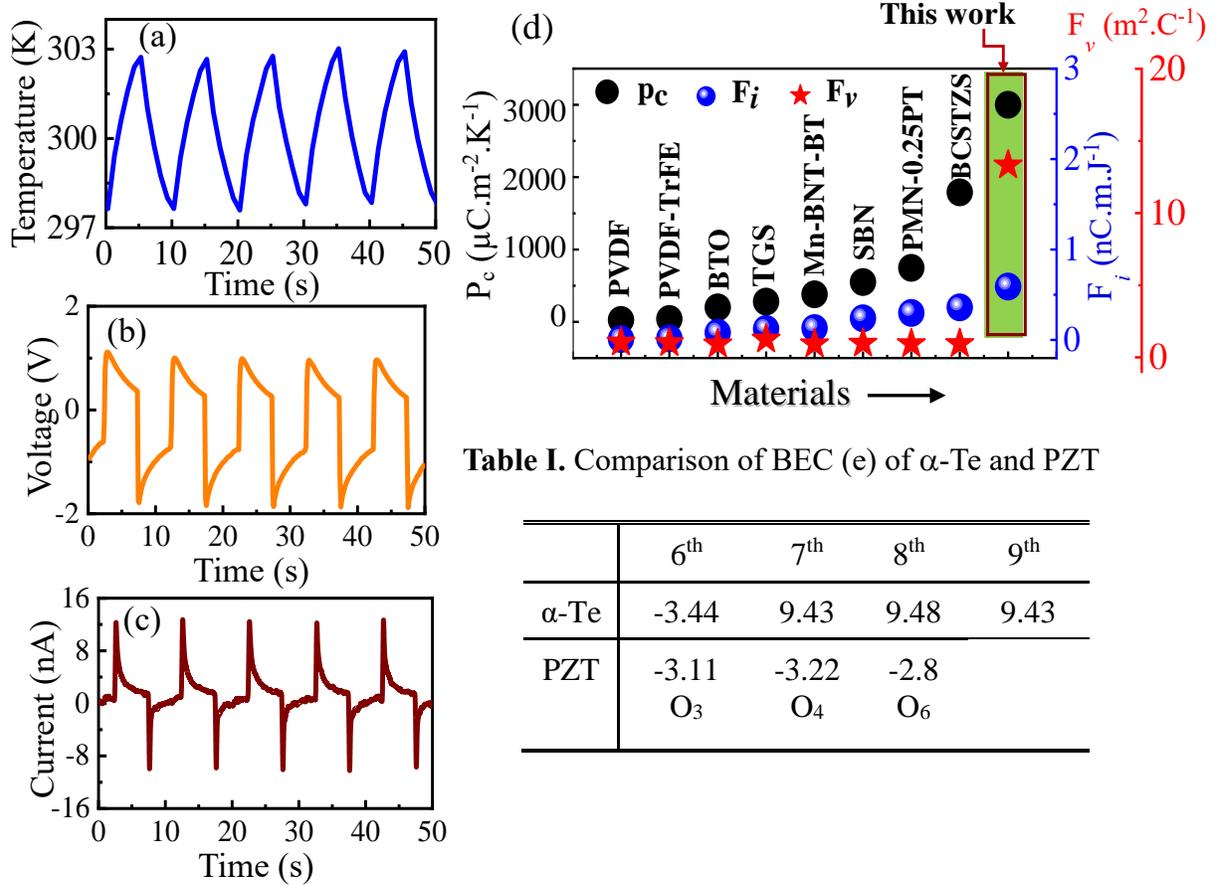

FIG. 4. (a) An acquired profile of a temperature at a 0.1 Hz frequency, corresponding outputs of (b) an open circuit voltage and (c) a short circuit current of the PyNG, and (d) a comparison of $P_c$, $F_i$ and $F_v$ values, with a giant pyroelectric response obtained for the 2D Te nanosheets in this work. Table 1 describes the BEC values.